\documentclass[twocolumn,showpacs,amsmath,amssymb,pra]{revtex4}

\usepackage{graphicx}   % Include figure files
\usepackage{dcolumn}    % Align table columns on decimal point
\usepackage{bm}         % bold math
\newcommand{\rd}{{\rm d}}

\newcommand{\kL}{k_{\rm L}}
\newcommand{\lL}{\lambda_{\rm L}}
\newcommand{\Er}{E_{\rm r}}
\begin{document}

\title[Multiphoton resonances]
      {ac Stark shift and multiphoton-like resonances in low-frequency driven 
      optical lattices}

\author{Stephan Arlinghaus}
\author{Martin Holthaus}

\affiliation{Institut f\"ur Physik, Carl von Ossietzky Universit\"at,
	D-26111 Oldenburg, Germany}

\date{May 15, 2012}

\begin{abstract}
We suggest that Bose-Einstein condensates in optical lattices subjected to 
ac forcing with a smooth envelope may provide detailed experimental access 
to multiphoton-like transitions between ac-Stark-shifted Bloch bands. Such 
transitions correspond to resonances described theoretically by avoided 
quasienergy crossings. We show that the width of such anticrossings can be 
inferred from measurements involving asymmetric pulses. We also introduce a 
pulse tracking strategy for locating the particular driving amplitudes for 
which resonances occur. Our numerical calculations refer to a currently 
existing experimental set-up [Haller {\em et al.\/}, PRL {\bf 104}, 200403 
(2010)].  
\end{abstract}

\pacs{03.75.Kk, 42.50.Hz, 67.85.Hj}

% 03.75.Kk 	Dynamic properties of condensates; collective and hydrodynamic 
%		excitations, superfluid flow
% 42.50.Hz 	Strong-field excitation of optical transitions in quantum 
%		systems; multiphoton processes; dynamic Stark shift
% 67.85.Hj 	Bose-Einstein condensates in optical potentials
 
\maketitle

%%%%%%%%%%%%%%%%%%%%%%%%%%%%%%%%%%%%%%%%%%%%%%%%%%%%%%%%%%%%%%%%%%%%%%%%%%%%%%%%

\section{Introduction}

The study of multiphoton excitation and ionization processes so far 
has concerned the interaction of matter with strong electromagnetic 
fields~\cite{MainfrayManus91,CohenTannoudjiEtAl92,DeloneKrainov00,ChuTelnov04}.    
Extending our previous works~\cite{ArlinghausHolthaus10,ArlinghausHolthaus11a}, 
we demonstrate in the present proposal that ideas and concepts developed in 
this field of research can also be applied for understanding the response of 
ultracold macroscopic matter waves in optical lattices to ac forcing with
a slowly varying envelope. Such systems offer additional handles of control 
which are not available in experiments with atoms or molecules exposed to 
pulses of laser radiation. Therefore, they can give novel insights into 
multiphoton dynamics in general, and in particular may also allow one to 
systematically investigate the effects of interparticle interactions on 
such transitions.    

The experimental investigation of Bose-Einstein condensates (BECs) in 
time-periodically forced optical lattices has been pushed forward with 
remarkable vigor within the last years, addressing quite diverse topics 
such as parametric amplification of matter waves~\cite{GemelkeEtAl05}, 
dynamic localization~\cite{LignierEtAl07,EckardtEtAl09}, 
photon-assisted tunneling~\cite{SiasEtAl08}, coherent control 
of the superfluid-to-Mott insulator transition~\cite{ZenesiniEtAl09}, 
quantum ratchets~\cite{SalgerEtAl09}, 
super Bloch oscillations~\cite{HallerEtAl10},
quantum simulation of frustrated magnetism~\cite{StruckEtAl11}, 
controlled correlated tunneling~\cite{MaEtAl11,ChenEtAl11},
and even the realization of tunable gauge potentials~\cite{StruckEtAl12}. 
Thus, BECs subjected to time-periodic forcing constitute an emerging branch
of research~\cite{ArimondoEtAl12}. Here we outline how systematic use of the
forcing's envelope will allow one to exctract crucial information about
the systems' dynamics.

\section{Detection of multiphoton-like resonances} 

\begin{figure}[t]
\includegraphics[width = 5.5cm]{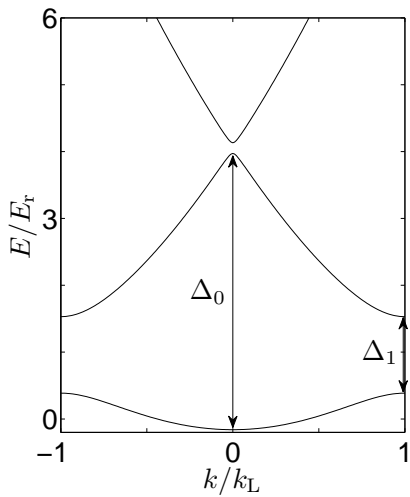}
\caption{Energy band structure of an unperturbed optical cosine lattice with 
    depth $V_0 = 2.3 \, \Er$. Measured in terms of the ``photon'' energy
    $\hbar\omega = 0.23 \, \Er$, as corresponding to the driving frequency
    $\omega/(2\pi) = 300$~Hz, the gap between the lowest two bands is
    $\Delta_1 = 5.05 \, \hbar\omega$ at the Brillouin zone edge, and
    $\Delta_0 = 18.24 \, \hbar\omega$ at its center.} 
\label{F_1}
\end{figure}

The present theoretical study of multiphoton-like condensate dynamics refers 
to conditions recently realized by Haller {\em et al.\/}~\cite{HallerEtAl10}: 
These authors have loaded BECs of Cs~atoms into a vertically oriented 1D 
optical lattice $V(x) = (V_0/2) \cos(2 \kL x)$. Here $\kL = 2\pi/\lL$, 
where $\lL = 1064.49$~nm is the wavelength of the lattice-generating laser 
light, so that the lattice period is $d = \pi/\kL$. We choose a comparatively 
shallow lattice with depth $V_0 = 2.3 \, \Er$, where $\Er = (\hbar \kL)^2/(2m)$
is the single-photon recoil energy~\cite{MorschOberthaler06}, with $m$ denoting
the mass of the atoms. The first excited band then is of the ``above-barrier'' 
type, so that atoms in this and all higher bands do not need to tunnel through 
the barriers when moving over the lattice. Using a magnetically induced 
Feshbach resonance, the $s$-wave scattering length of the Cs atoms is tuned 
to zero~\cite{KohlerEtAl06}, so that one is dealing with a condensate of 
effectively noninteracting particles. By means of a time-periodic modulation 
of the levitation gradient employed for compensating gravity, an external 
oscillating force is introduced which acts on the atoms with maximum amplitude 
$F_{\rm max}$ and frequency $\nu = \omega/(2\pi)$, such that their dynamics 
are governed by the Hamiltonian  
\begin{equation}
	H(t) = \frac{p^2}{2m} + V(x) - s(t) F_{\rm max} x \cos(\omega t)
\label{eq:HAM}
\end{equation}
with a dimensionless shape function $s(t)$ determining the envelope of the
pulses applied. All calculations reported here are performed for the relatively
low frequency $\omega/(2\pi) = 300$~Hz, so that $\hbar\omega = 0.23 \, \Er$. As 
indicated in Fig.~\ref{F_1}, this means that the gap $\Delta_1 = 1.14 \, \Er$ 
between the lowest two Bloch bands $E_1(k)$ and $E_2(k)$ of the unperturbed 
optical lattice, which occurs at the Brillouin zone edge $k/\kL = \pm 1$, 
amounts to $5.05 \, \hbar\omega$: Exciting particles from the initially 
occupied lowest band to the first excited one requires the absorption of more 
than five ``photons''.

A first glimpse at the condensate dynamics under the action of a force 
$F(t)$ is provided by Bloch's acceleration theorem: The expectation value 
$\langle k \rangle(t)$ of an atomic wave packet in $k$-space evolves in time 
according to 
\begin{equation}
	\hbar \frac{\rd}{\rd t} \langle k \rangle(t) = F(t) \; ,
\label{eq:ACT}
\end{equation} 
provided interband transitions remain negligible~\cite{Bloch29}. Assuming 
a sinusoidal force $F(t) = F \cos(\omega t)\Theta(t)$ instantaneously 
switched on at time $t = 0$, this gives $\langle k \rangle(t)/\kL = 
\langle k \rangle(0)/\kL + (K/\pi)\sin(\omega t)$, where $K = Fd/(\hbar\omega)$
is a dimensionless measure of the driving amplitude. Thus, when considering
a wave packet initially at rest in the lowest band, so that 
$\langle k \rangle(0)/\kL = 0$, the packet's center $\langle k \rangle(t)$ 
just reaches the Brillouin zone edge when $K = \pi$. Because the zone edge 
gives rise to Zener-type transitions~\cite{Zener34}, one then expects 
pronounced excitation of higher bands. Therefore, one has $K \approx \pi$ 
as a rough order-of-magnitude estimate of the amplitude required for inducing 
multiphoton-like transitions.

\begin{figure}
\includegraphics[width = 7cm]{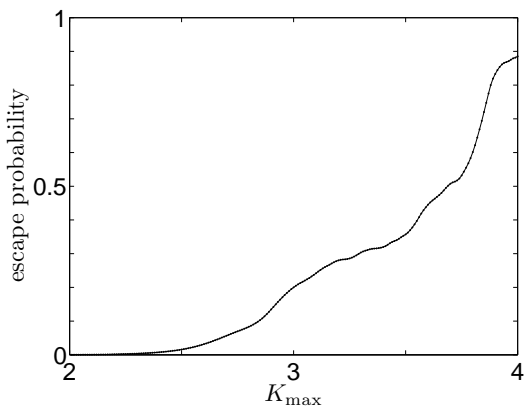}
\caption{Interband transition probabilities after pulses with the squared-sine 
	envelope~(\ref{eq:ENV}) and length $T_{\rm p} = 60 \times 2\pi/\omega$,
	obtained from numerical solutions of the Schr\"odinger equation. All 
	atoms which are excited to bands $n > 1$ after a pulse are assumed to 
	escape from the lattice.}   	 
\label{F_2}
\end{figure}

This expectation is confirmed by the numerical calculations summarized 
in Fig.~\ref{F_2}. Here we consider driving pulses 
$s(t) F_{\rm max}\cos(\omega t)$ as already incorporated into the 
Hamiltonian~(\ref{eq:HAM}), with a smooth envelope
\begin{equation}
	s(t) = \sin^2(\pi t / T_{\rm p}) 
	\quad ; \quad 
	0 \le t \le T_{\rm p} \; .
\label{eq:ENV}
\end{equation}
The pulse length is fixed at 60 driving cycles, $T_{\rm p} = 60 \, T$ 
with $T = 2\pi/\omega$. In order to model the dynamics of a noninteracting
BEC responding to such pulses, we start with an initial state
\begin{equation}
  	\psi(x, 0) = \sqrt{\frac{d}{2\pi}} \int \! \rd k \, 
  	g_1(k,0) \chi_{1,k}(x) 
\label{eq:IST}
\end{equation}
made up from Bloch waves $\chi_{1,k}(x)$ of the lowest band, employing a 
Gaussian momentum distribution
\begin{equation}
  	g_1(k,0) = \left(\sqrt{\pi}\Delta k\right)^{-1/2}  
	\exp\!\left(-\frac{[k - \langle k \rangle(0)]^2}
		{2\left(\Delta k\right)^2}\right)
\label{eq:IMD}
\end{equation}
centered around $\langle k \rangle(0)/\kL = 0$ with width $\Delta k/\kL = 0.1$,
and then solve the single-particle Schr\"odinger equation by means of a 
Crank-Nicolson algorithm~\cite{CrankNicolson47}. Varying the maximum scaled 
amplitude $K_{\rm max} = F_{\rm max}d/(\hbar\omega)$ from pulse to pulse, we 
plot the escape probability from the lattice at the end of each pulse, at 
$t = T_{\rm p}$. Here we assume that only atoms which finally still populate 
the lowest band remain in the lattice, since atoms which have been excited to 
the higher ``above-barrier'' bands tend to escape from the shallow lattice 
quite fast. Evidently, interband transitions start to make themselves felt at  
$K_{\rm max} \approx 2.5$, and reach substantial strength when 
$K_{\rm max} \approx 3$, confirming the above rough estimate. Thus, 
proof-of-principle experiments performed along these lines should establish 
the feasibility of using BECs in driven optical lattices as novel probes for 
multiphoton-like transitions: Take driving pulses with smooth envelopes, and
measure the interband transition probability at the end of each pulse. In 
a series of such measurements with a constant pulse shape one then should 
observe a pronounced onset of interband transitions when the maximum amplitude
is successively increased.

\section{Avoided-quasienergy-crossing spectroscopy with asymmetric pulses}

\begin{figure}[t]
\includegraphics[width = 7cm]{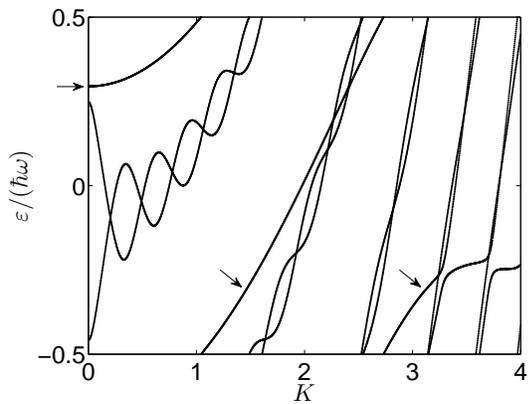}
\caption{Quasienergies $\varepsilon_n(0)$ at the center of the Brillouin zone
	for an optical lattice with depth $V_0/\Er = 2.3$, driven with scaled 
	frequency $\hbar\omega/\Er = 0.23$. The quasienergy originating from 
	the lowest band $n = 1$ is marked by arrows. It exhibits a 
	substantial ac Stark shift, and undergoes pronounced avoided crossings 
	for $K > 3$.}   	 
\label{F_3}
\end{figure}

In a second step, this approach can be employed for getting more detailed 
insight into multiphoton dynamics. Namely, the single-band acceleration 
theorem~(\ref{eq:ACT}) ignores an essential element: Not only does the wave
packet's center $\langle k \rangle(t)$ move within its band in response to 
the external forcing, but also the bands themselves are ``dressed'' by the 
drive, and therefore experience an ac Stark shift~\cite{ArlinghausHolthaus11b}. 
Hence, initially nonresonant bands may be shifted such that their separation 
in energy approaches an integer multiple of $\hbar\omega$ for certain Bloch 
wavenumbers~$k$, possibly leading to strong resonant interband coupling, that 
is, to a multi\-photon resonance. Theoretically, such resonances are found 
by computing the quasienergy bands $\varepsilon_n(k)$ which emerge from the 
energy bands $E_n(k)$ in the presence of a drive with constant amplitude. These
quasienergy bands reflect the ac-Stark-shifted energy bands, projected into 
an interval of width $\Delta\varepsilon = \hbar\omega$, so that a multiphoton 
resonance translates into an avoided quasienergy crossing~\cite{ChuTelnov04,
ArlinghausHolthaus10,ArlinghausHolthaus11a}. For example, Fig.~\ref{F_3} shows 
the quasienergies $\varepsilon_n(0)$ with $n = 1,2,3$, which pertain to the 
pulses considered in Fig.~\ref{F_2}, plotted versus the scaled amplitude~$K$. 
The quasienergy originating from the ground-state energy $E_1(0)$ of the 
optical lattice undergoes two well-resolved avoided crossings when $K > 3$, 
signaling the presence of two individual multiphoton resonances. The 
observation that these ``large'' resonances begin to show up only for 
$K \approx \pi$ nicely relates the elaborate quasienergy approach to the 
previous elementary reasoning based on Eq.~(\ref{eq:ACT}).

\begin{figure}
\includegraphics[width = 7cm]{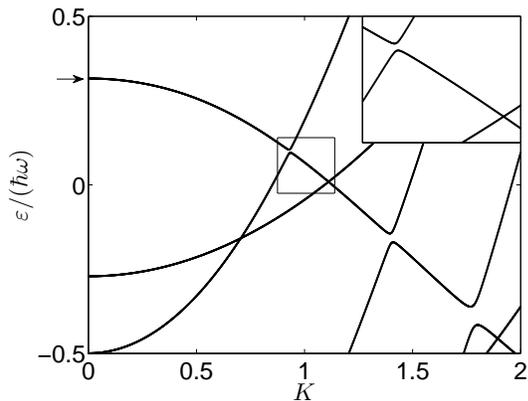}
\caption{Quasienergies $\varepsilon_n(k)$ for $k/\kL = 0.8$ and $n = 1,2,3$.
	The arrow indicates the quasienergy originating from the lowest
	energy band $n = 1$.}     	 
\label{F_4}
\end{figure}

Inspecting that elementary reasoning once again, one expects multiphoton
resonances to occur for smaller driving amplitudes when the initial packet is 
centered around a nonzero wavenumber, $\langle k \rangle(0)/\kL \neq 0$, since
then smaller amplitudes are required for reaching the Brillouin zone edge. 
This expectation is confirmed by Fig.~\ref{F_4}, which depicts quasienergies 
$\varepsilon_n(k)$ for $k/\kL = 0.8$, with the first resonance showing up 
already at $K \approx 0.9$. Experimentally, one can prepare an initial state 
with arbitrary $\langle k \rangle(0)$ by subjecting the condensate to a 
suitable ``kick''~\cite{BurgerEtAl01,DenschlagEtAl02}. Thus, one should also 
be able to detect the resonances predicted by Fig.~\ref{F_4}. To this end, we 
propose a particular kind of ``avoided-quasienergy-crossing spectroscopy'' 
based on the use of {\em asymmetric\/} pulses $s(t)$. For illustration, we 
assume that the rising part of such pulses be given by the first half of the  
envelope~(\ref{eq:ENV}), with fixed switch-on time $T_{\rm p}^{(1)}/2 = 5 \, T$,
while their decreasing part is described by the second half of a squared-sine 
envelope, but with a different switch-off time $T_{\rm p}^{(2)}/2$. 
Let the maximum scaled amplitude be $K_{\rm max} = 1.2$. During the rising 
part of such a pulse, a wave packet initially centered around 
$\langle k \rangle(0)/\kL = 0.8$ then follows its quasienergy states 
adiabatically, until the instantaneous amplitude reaches the multiphoton 
resonance at $K \approx 0.9$ visible in Fig.~\ref{F_4}. Then the packet 
undergoes a Landau-Zener transition to the anticrossing quasienergy 
state~\cite{DreseHolthaus99}. Due to the rapid switch-on of the pulse, and 
to the narrow quasienergy separation $\delta \varepsilon$ at the avoided 
crossing, that transition is almost complete. Thereafter, the packet again 
adiabatically follows the pulse envelope, until the resonance is encountered 
a second time when the  amplitude decreases. If then 
$T_{\rm p}^{(2)} \gg T_{\rm p}^{(1)}$, a major part of the wave function does 
not ``jump over'' the avoided crossing back to the initial state, but rather 
stays in the continuously connected quasienergy states. This implies that a 
major fraction of the condensate atoms is excited at the end of the pulse, 
escaping out of the lattice. When such an experiment is performed repeatedly 
with fixed rise time $T_{\rm p}^{(1)}/2$ while varying the switch-off duration 
$T_{\rm p}^{(2)}/2$, one should observe survival probabilities which drop 
exponentially with increasing $T_{\rm p}^{(2)}$, allowing one to extract the 
quasienergy separation $\delta\varepsilon$ at the avoided crossing from the 
drop rate by means of the known Landau-Zener formula for quasienergy 
states~\cite{DreseHolthaus99}.

\begin{figure}[t]
\includegraphics[width = 7cm]{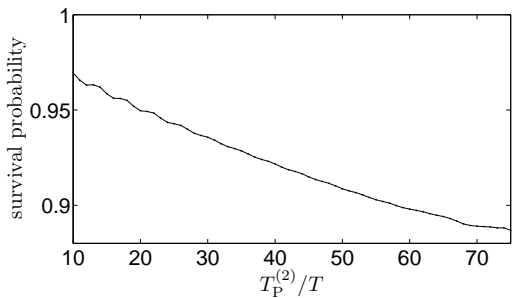}
\caption{Survival probability of atoms in the lowest band after asymmetric
	pulses with fast switch-on time $T_{\rm p}^{(1)}/2 = 5 \, T$, and
	with  varying switch-off durations $T_{\rm p}^{(2)}/2$. The initial 
	packets were centered around $\langle k \rangle(0)/\kL = 0.8$, so that 
	the dynamics are determined by the spectrum shown in Fig.~\ref{F_4};
	$K_{\rm max} = 1.2$ for all pulses.}
\label{F_5}
\end{figure}

The results of a series of solutions to the Schr\"odinger equation 
corresponding to this scenario are plotted in Fig.~\ref{F_5}. From the
slope of the numerical data we deduce a quasienergy gap 
$\delta\varepsilon/(\hbar\omega) = 0.0099$, in agreement with the value $0.01$ 
read off from Fig.~\ref{F_4}. These findings clearly underline that this 
quasienergy gap $\delta\varepsilon \ll \hbar\omega$ is the actually relevant 
energy scale for the multiphoton transitions under scrutiny here, {\em not\/} 
the band separation $\Delta_1 = 5.05 \, \hbar \omega$ indicated in 
Fig.~\ref{F_1}.

\begin{figure}[t]
\includegraphics[width = 7cm]{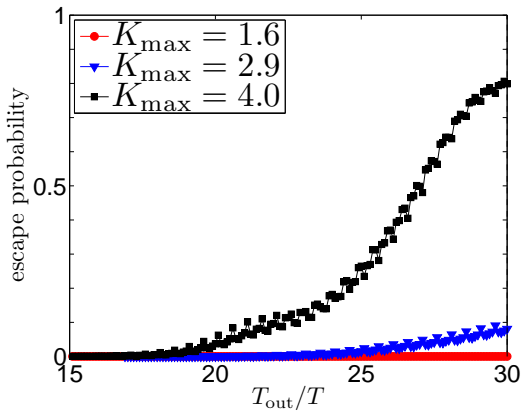}
\caption{(color online) Pulse tracking realized by switching off the squared-sine 
	envelope~(\ref{eq:ENV}) with $T_{\rm p} = 60 \, T$ abruptly at 
	$t = T_{\rm out}$, and recording the escape probability at this 
	moment. The onset of interband transitions then reveals the presence 
	of a multiphoton resonance at the instantaneous amplitude reached 
	at $t = T_{\rm out}$.}     
\label{F_6}
\end{figure}

There are further options offered by BECs in driven optical lattices which have
no match in laser-based multiphoton studies. For example, one can switch off 
the driving force abruptly at any moment, and analyze the state of the wave 
packet at that particular instant. Specifically, we again utilize envelopes 
of the form~(\ref{eq:ENV}), and let the amplitude rapidly drop to zero at 
some instant $T_{\rm out}$, with $0 < T_{\rm out} < T_{\rm p}/2$. We set 
$T_{\rm p} = 60 \, T$ and take initial wave packets with 
$\langle k \rangle(0)/\kL = 0$, as for the previous calculations shown 
in Fig.~\ref{F_2}, and plot the escape probabilities versus switch-off time 
$T_{\rm out}$ in Fig.~\ref{F_6}. For $K_{\rm max} = 1.6$ the wave packet 
simply follows its quasienergy states adiabatically, not encountering any of 
the resonances observed in Fig.~\ref{F_3}. However, for $K_{\rm max} = 2.9$ 
interband transitions set in at $T_{\rm out} \approx 25 \, T$, that is, 
when the instantaneous amplitude reaches 
$K = K_{\rm max} s(T_{\rm out}) \approx 2.7$; this is due to a tiny resonance 
not resolved in Fig.~\ref{F_3}. When $K_{\rm max} = 4.0$, pronounced interband 
transitions occur at $T_{\rm out} \approx 20 \, T$, corresponding to the 
instantaneous amplitude $K \approx 3$: This is already in the regime of 
influence of the first of the two ``large'' resonances seen in Fig.~\ref{F_3}. 
Thus, this ``pulse tracking'' strategy allows one to experimentally detect 
multiphoton resonances, that is, to find those values of the driving amplitude 
for which strong resonant interband transitions occur.\\

\section{Summary and outlook}

To summarize, we have argued that BECs in shallow optical lattices exposed to 
ac forcing with frequencies of some 100~Hz and smooth envelopes can be employed
for mimicking multiphoton processes. The enormous degree of controllability 
realizable with such set-ups enables one to obtain information not reachable 
with laser-irradiated crystalline solids~\cite{ArlinghausHolthaus10}; in 
particular, we have suggested the use of asymmetric pulses for performing 
avoided-quasienergy-crossing spectroscopy. Moreover, we have shown how pulse 
tracking by abruptly switching off the driving amplitude allows one to monitor 
the dynamics at each moment during an individual pulse, and thus to locate 
multiphoton resonances between ac-Stark-shifted Bloch bands. Obviously, 
our approach also lends itself to systematic explorations of the effects of 
pulse shaping. Even more interestingly, one can activate interparticle 
interactions by suitably tuning the $s$-wave scattering length of the Cs 
atoms~\cite{KohlerEtAl06}. Hence, the detailed experimental investigation of 
the influence of such interactions on multiphoton transitions has come into 
immediate reach.

\begin{acknowledgments}
We are indebted to E.~Haller for detailed discussions, and for performing
preliminary measurements confirming the viability of our suggestions. We
also acknowledge support from the Deutsche Forschungsgemeinschaft under 
grant No.\ HO 1771/6.
\end{acknowledgments}

\end{document}